\documentclass[journal]{IEEEtran}

\usepackage{amsmath}
\usepackage{mathtools}
\usepackage{amssymb}
\ifCLASSINFOpdf
\else
   \usepackage[dvips]{graphicx}
\fi
\usepackage{url}
\usepackage[noadjust]{cite}
\usepackage{multirow}

\usepackage{graphicx}
\usepackage{array}
\newcolumntype{?}{!{\vrule width 1.25pt}}
\newcolumntype{@}{!{\vrule width 1.6pt}}

\newcommand{\mycomment}[1]{}
\begin{document}

\title{Towards Maximum Likelihood Training for Transducer-based Streaming Speech Recognition}

\author{Hyeonseung Lee, Ji Won Yoon, Sungsoo Kim, and Nam Soo Kim \IEEEmembership{Senior Member, IEEE}
\thanks{This work was supported by the COMPA grant funded by the Korea government (MSIT and Police) (No. RS2023-00235082) and the Institute of Information \& communications Technology Planning \& Evaluation (IITP) grant funded by the Korea government (MSIT) (2021-0-01341, Artificial Intelligence Graduate School Program (Chung-Ang University)).
}
\thanks{Hyeonseung Lee is with XL8 Inc. (e-mail: swigls1@gmail.com). Sungsoo Kim and Nam Soo Kim are with the Institute of New Media and Communications, Department of Electrical and Computer Engineering, Seoul National University, Seoul, Republic of Korea (sskim@hi.snu.ac.kr; nkim@snu.ac.kr). Ji Won Yoon is with the Department of Artificial Intelligence, Chung-Ang University, Seoul, Republic of Korea (jiwonyoon@cau.ac.kr). 

© 2024 IEEE.  Personal use of this material is permitted.  Permission from IEEE must be obtained for all other uses, in any current or future media, including reprinting/republishing this material for advertising or promotional purposes, creating new collective works, for resale or redistribution to servers or lists, or reuse of any copyrighted component of this work in other works.

Digital Object Identifier 10.1109/LSP.2024.3491019
}}

\markboth{IEEE Signal Processing Letters, Vol. 27, 2024}
{Shell \MakeLowercase{\textit{et al.}}: Bare Demo of IEEEtran.cls for IEEE Journals}
\maketitle

\begin{abstract}
Transducer neural networks have emerged as the mainstream approach for streaming automatic speech recognition (ASR), offering state-of-the-art performance in balancing accuracy and latency.
In the conventional framework, streaming transducer models are trained to maximize the likelihood function based on non-streaming recursion rules. However, this approach leads to a mismatch between training and inference, resulting in the issue of deformed likelihood and consequently suboptimal ASR accuracy.
We introduce a mathematical quantification of the gap between the actual likelihood and the deformed likelihood, namely forward variable causal compensation (FoCC). We also present its estimator, FoCCE, as a solution to estimate the exact likelihood.
Through experiments on the LibriSpeech dataset, we show that FoCCE training improves the accuracy of the streaming transducers.
\end{abstract}


\IEEEpeerreviewmaketitle

\vspace{-0.2in}
\section{Introduction}

\IEEEPARstart{M}{odern} automatic speech recognition (ASR) has witnessed substantial accuracy improvements, primarily attributed to advances in deep learning. While the pursuit of higher accuracy in general ASR remains a priority, recent studies have increasingly emphasized the need to maintain accuracy in challenging scenarios, including spoken named entities~\cite{peyser2020improving_proper_noun, sim2019personalization_named_entity}, multi-lingual speech~\cite{zhang2023google_usm, pratap2023scaling_mms}, and streaming ASR~\cite{li2021better_streaming, narayanan2021cascaded, yu2021dual_mode_ASR}. Notably, the rising demand for on-device and real-time ASR underscores the importance of streaming ASR.

The accuracy of streaming ASR is degraded compared to its non-streaming counterpart, especially when the model is restricted to have low latency.
Two separate causes induce the accuracy degradation of the streaming model.
The primary cause is \textit{information deficiency};
the streaming model predicts an output based on a limited input context, whilst the non-streaming model has the advantage of being aware of the entire context.
Information deficiency is an inevitable intrinsic property of streaming ASR.
The second cause is \textit{deformed likelihood}.
Deep learning ASR models are usually trained based on the maximum likelihood criterion \cite{hinton2012deep, CTC, RNN-T, Bahdanau, LAS, dong2019cif, xiang2019crf, chan2020imputer, variani2022global}.
Mainstream ASR approaches~\cite{CTC, RNN-T, LAS} commonly define likelihood by breaking it into local probability terms that are estimated using neural networks with a softmax output layer.
These local probabilities are designed to sum up to the likelihood when all local probabilities are modeled with the entire input context, which is the case for non-streaming ASR but not valid for streaming ASR~\cite{variani2022global}.
Consequently, naively utilizing mainstream ASR likelihood functions for training the streaming models leads to the deformed likelihood problem.

Several approaches~\cite{xiang2019crf, variani2022global} adopt a globally normalized likelihood definition, which bypasses the issue of deformed likelihood in streaming ASR.
These methods define globally normalized likelihood as a ratio between the score-sum of accepting paths (corresponding to ground truth data) and the score-sum of all possible paths.
By excluding local probability terms, these approaches avoid the deformed likelihood problem.
However, they exhibit lower accuracy compared to local probability-based mainstream ASR methods in both streaming and non-streaming scenarios. This limitation restricts their application, as globally normalized likelihood is incompatible with local probability-based models.

Transducer neural networks \cite{RNN-T}, often called RNN-T, have recently emerged as the dominant approach for streaming ASR, offering a state-of-the-art tradeoff between accuracy and latency.
The prevalent paradigm in streaming transducer models relies on a likelihood function derived from the non-streaming recursion rule.
Training a transducer with this naive likelihood induces the deformed likelihood problem, resulting in sub-optimal ASR accuracy.

This letter introduces a mathematical perspective on the deformed likelihood problem in streaming transducer training and proposes a novel solution to mitigate it.
The key contributions of this paper are as follows:
\begin{itemize}
\item We reframe the dynamic programming for the non-streaming transducer likelihood~\cite{RNN-T} using detailed probabilistic notation and demonstrate its mismatch to a streaming model due to the \textit{deformed likelihood} problem.
\item We quantify the gap between the deformed and the actual likelihood in streaming transducer training, namely ``\textbf{Fo}rward Variable \textbf{C}ausal \textbf{C}ompensation" (FoCC).
\item We propose the FoCC estimator (FoCCE) network, which estimates the actual likelihood instead of the deformed likelihood in streaming transducer training.
\item We experimentally show that FoCCE training improves the streaming transducers' ASR accuracy on the LibriSpeech dataset, reducing the accuracy gap between streaming and non-streaming transducers.
\end{itemize}

\section{Transducer Neural Networks}
\label{sec:transducer}

\subsection{Non-streaming Transducer}

Given an input sequence $\mathbf{x}_{1:T}$ of length $T$
and the corresponding target sequence $\mathbf{y}_{0:U}$ of length $U$ ($y_u{\in}\mathbb{N}$, $y_0$ is the start-of-sequence token $\langle sos \rangle$), a transducer neural network~\cite{RNN-T} parametrized by $\theta$ computes the conditional likelihood 
$L_{\theta} (\mathbf{x}_{1:T},\mathbf{y}_{0:U})$ as
\begin{equation} 
\begin{split}
\label{eq:conditional_likelihood}
L_{\theta} (\mathbf{x}_{1:T},\mathbf{y}_{0:U})
    := \log P_\theta (\mathbf{y}_{0:U}, z_{U}\leq T<z_{U+1} | \mathbf{x}_{1:T}),
\end{split} 
\end{equation}
\begin{equation} 
\label{eq:encoder}
\mathbf{f}_{1:T} = \mathrm{Encoder}_\theta (\mathbf{x}_{1:T})
\end{equation}
where the latent alignment variable $z_{u}\in\mathbb{N}$ is defined such that $z_u=t$ means that the target $y_u$ is aligned to the encoded input $f_t$. 
$\mathrm{Encoder}_\theta (\cdot)$ is a neural network that extracts abstract 
information from the entire input sequence.
The encoder may subsample its input,
i.e., $\mathbf{f}_{1:T} = \mathrm{Encoder}_\theta (\mathbf{x}_{1:T'})$ where $T < T'$, to reduce the length mismatch between input and target.
As any long input sequence can be reshaped into a chunked sequence that has the same length as the encoded sequence, i.e., ${x}_{1:T'}={x'}_{1:T}$, we denote the input as ${x}_{1:T}$ for simplicity.

Since the alignments between the inputs and targets are assumed to be monotonic, $\mathbf{z}_{0:U}$ is constrained such that
\begin{equation} 
\label{eq:monotonic_constraint}
z_0 \leq 1 \leq z_1 \leq z_2 \leq z_3 \leq \dots \leq z_U.
\end{equation}
The end condition $z_{U}\leq T<z_{U+1}$ in \eqref{eq:conditional_likelihood} indicates that the entire target $\mathbf{y}_{1:U}$ is aligned to the input $\mathbf{x}_{1:T}$ whilst the next target $y_{U+1}$ is not (i.e., only $\mathbf{y}_{0:U}$ is included in the inputs).

Dynamic programming is used to obtain $L_{\theta} (\mathbf{x}_{1:T},\mathbf{y}_{0:U})$ in $\mathcal{O}(TU)$ computations. The forward variable is defined as 
\begin{equation} 
\label{eq:forward_variable}
\alpha_\theta (t,u):= P_\theta (\mathbf{y}_{0:u}, z_u \leq t \leq z_{u+1} | \mathbf{x}_{1:T} ).
\end{equation}
From \eqref{eq:monotonic_constraint} and \eqref{eq:forward_variable}, at the initial point $\alpha_\theta (1,0) = P_\theta (y_0=\langle sos \rangle |\mathbf{x}_{1:T}) = 1$ and at the boundaries $\alpha_\theta (t,-1)=\alpha_\theta (0,u)=0$ for $t \in [1,T]$ and $u \in [0,U]$.
Local probabilities 
in $x$-axis and $y$-axis are respectively defined as
\begin{equation} 
\begin{split}
\phi_\theta (t,u):= P_\theta (\ \: \,  z_{u+1} \geq t+1|& \mathbf{x}_{1:T}, \mathbf{y}_{0:u}, z_u \leq t \leq z_{u+1}), \\
Y_\theta (t,u):= P_\theta (y_{u+1}, z_{u+1}=t|& \mathbf{x}_{1:T}, \mathbf{y}_{0:u}, z_u \leq t \leq z_{u+1}),    
\end{split}
\end{equation}
for $t \in [1,T], u \in [0,U]$, which is to be estimated by neural networks.
With blank probability $\phi_\theta (\cdot,\cdot)$ and label probability $Y_\theta (\cdot,\cdot)$, the forward variable can be recursively computed as
\begin{equation} 
\begin{split}
\label{eq:transducer_recursion}
\alpha_\theta (t,u)=&\quad \: P_\theta (\mathbf{y}_{0:u}, z_u < t \leq z_{u+1} | \mathbf{x}_{1:T} ) \\
        &+ P_\theta (\mathbf{y}_{0:u}, z_u = t \leq z_{u+1} | \mathbf{x}_{1:T} ) \\
        =& \ \; P_\theta (\mathbf{y}_{0:u}, z_u \leq t-1 \leq z_{u+1} | \mathbf{x}_{1:T} ) \\
        & \cdot P_\theta (z_{u+1} \geq t | \mathbf{x}_{1:T}, \mathbf{y}_{0:u}, z_{u} \leq t-1 \leq z_{u+1}) \\
        &+ P_\theta (\mathbf{y}_{0:u-1}, z_{u-1} \leq t \leq z_{u} | \mathbf{x}_{1:T} ) \\
        &\ \, \cdot P_\theta (y_{u}, z_{u} = t | \mathbf{x}_{1:T}, \mathbf{y}_{0:u-1}, z_{u-1} \leq t \leq z_u) \\
        =& \alpha_\theta (t-1,u) \phi_\theta (t-1,u) + \alpha_\theta (t,u-1) Y_\theta (t,u-1) ,
\end{split}
\end{equation}
according to \eqref{eq:monotonic_constraint}.
The conditional likelihood in \eqref{eq:conditional_likelihood} can be obtained by
\begin{equation} 
\label{eq:conditional_likelihood_forward}
L_{\theta} (\mathbf{x}_{1:T},\mathbf{y}_{0:U})
    = \log \alpha_\theta (T,U) \phi_\theta (T,U),
\end{equation} 
which is used as a training 
objective.

In this subsection, we introduce the alignment variable $z_u$ separately from the target label variable $y_u$.
This separation clarifies the Bayes' rule within the transducer recursion, as shown in \eqref{eq:transducer_recursion}.
We found that the forward-backward algorithm~\cite{RNN-T}, which is a training method for transducer networks in most prior studies, does not obey Bayes' rule.
Therefore this paper focuses on the training based on the forward variable recursion rather than the forward-backward algorithm.


\subsection{Streaming Transducer}

\begin{figure*}[t]
\centerline{\includegraphics[width=\linewidth]{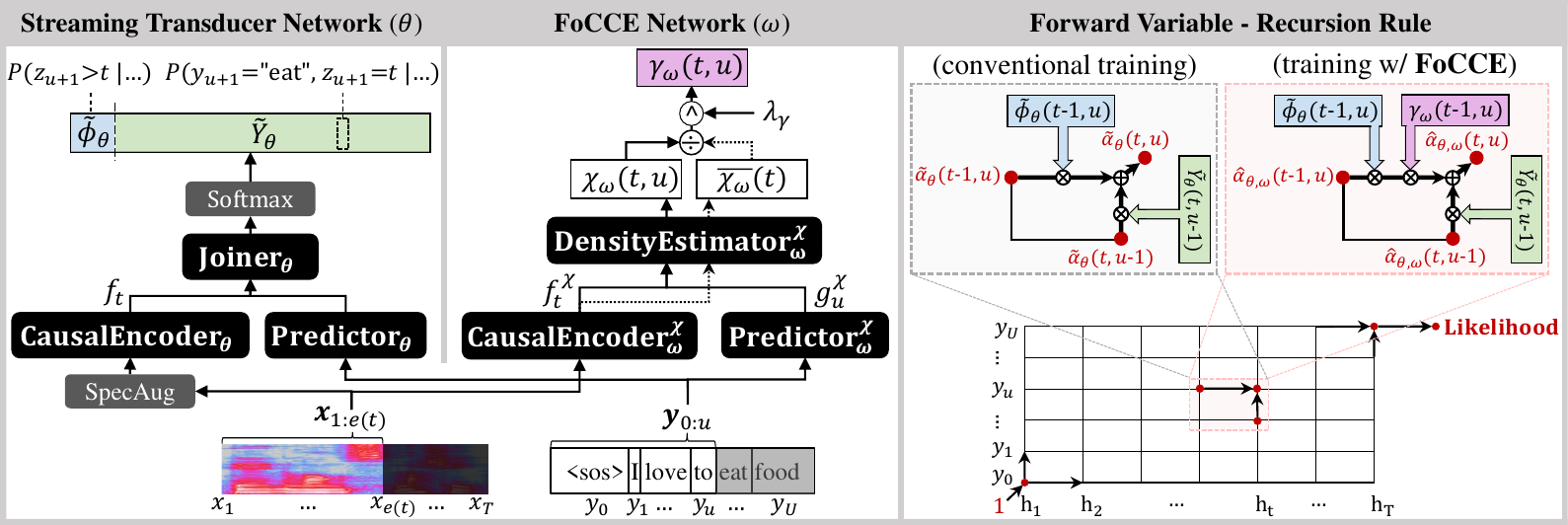}}
\caption{An illustration of the proposed FoCCE training.
The streaming transducer network (left) and the FoCCE network (middle) respectively estimate the local probabilities and FoCC values, which are used to estimate the actual likelihood by the modified forward variable recursion rule (red boxes on the right).}
\label{fig:pictorial_description}
\vspace{-0.15in}
\end{figure*}

In the streaming case, as depicted on the left of Figure~\ref{fig:pictorial_description}, a transducer network $\theta$ estimates the local probabilities at the $t$-th timestep using only limited input context $\mathbf{x}_{1:e(t)}$:
\begin{equation} 
\label{eq:streaming_encoder}
\mathbf{f}_{1:t} = \mathrm{CausalEncoder}_\theta (\mathbf{x}_{1:e(t)}),
\end{equation}
\begin{equation} 
\label{eq:streaming_phi_Y}
\begin{split}
\Tilde{\phi}_\theta (t,u):= P_\theta (\ \: \,  z_{u+1} \geq t+1|& \mathbf{x}_{1:e(t)}, \mathbf{y}_{0:u}, z_u \leq t \leq z_{u+1}), \\
\Tilde{Y}_\theta (t,u):= P_\theta (y_{u+1}, z_{u+1}=t|& \mathbf{x}_{1:e(t)}, \mathbf{y}_{0:u}, z_u \leq t \leq z_{u+1}),
\end{split}
\end{equation}
where the $\mathrm{CausalEncoder}_\theta (\cdot)$ is similar to the $\mathrm{Encoder}_\theta (\cdot)$ in \eqref{eq:encoder}, except that its input context is limited by a context-end function $e: \mathbb{N} \rightarrow \mathbb{N} $.
In general, $e(t)= \min (T, C \lceil t / C \rceil + R)$ with chunk size $C$ and right context offset $R$.

Conventional methods train streaming transducer models by maximizing the likelihood given by \eqref{eq:conditional_likelihood_forward}, with local probabilities in \eqref{eq:transducer_recursion} being substituted by \eqref{eq:streaming_phi_Y}.
This naive approach breaks Bayes' rule, thereby training models with deformed likelihood.
To obtain the actual likelihood, we introduce a probability ratio, namely \textbf{Fo}rward Variable \textbf{C}ausal \textbf{C}ompensation (FoCC):
\begin{equation}
\begin{split}
\label{eq:FoCC}
\gamma_\theta (t,u):=& \frac{P_\theta (\mathbf{y}_{0:u}, z_u \leq t < z_{u+1} | \mathbf{x}_{1:e(t+1)} ) \hfill}{P_\theta (\mathbf{y}_{0:u}, z_u \leq t < z_{u+1} | \mathbf{x}_{1:e(t)} ) \hfill} \\
    =& \frac{P_\theta (\mathbf{x}_{e(t)+1:e(t+1)} |\mathbf{x}_{1:e(t)}, \mathbf{y}_{0:u}, z_u \leq t < z_{u+1} )}{P_\theta (\mathbf{x}_{e(t)+1:e(t+1)} |\mathbf{x}_{1:e(t)}) \hfill}.
\end{split}
\end{equation}
Note that 
if $t$ and $t+1$ are in the same encoder chunk, i.e., $e(t)=e(t+1)$, then $\gamma_\theta (t,\cdot)=1$ as shown in \eqref{eq:FoCC}. 
Thus, FoCC needs to be calculated only for $t: e(t)<e(t+1)$.

The modified recursion rule can be formulated with the streaming forward variable, which is defined as
\begin{equation}
\begin{split}
\label{eq:streaming_forward_variable_and_transducer_recursion}
\Tilde{\alpha}_\theta (t,u):=& P_\theta (\mathbf{y}_{0:u}, z_u \leq t \leq z_{u+1} | \mathbf{x}_{1:e(t)} ) \\
        =& \quad \:  P_\theta (\mathbf{y}_{0:u}, z_u < t \leq z_{u+1} | \mathbf{x}_{1:e(t-1)} ) \gamma_\theta (t-1,u) \\
        &+ P_\theta (\mathbf{y}_{0:u}, z_u = t \leq z_{u+1} | \mathbf{x}_{1:e(t)} ) \\    
        =& \quad \:   \Tilde{\alpha}_\theta (t-1,u) \Tilde{\phi}_\theta (t-1,u) \gamma_\theta (t-1,u)\\
        &+ \Tilde{\alpha}_\theta (t,u-1) \Tilde{Y}_\theta (t,u-1).
\end{split}
\end{equation}
According to \eqref{eq:conditional_likelihood},~\eqref{eq:streaming_phi_Y}, and \eqref{eq:streaming_forward_variable_and_transducer_recursion}, the actual likelihood is
\begin{equation}
\label{eq:condition_likelihood_forward_streaming}
L_{\theta} (\mathbf{x}_{1:T},\mathbf{y}_{0:U})
    = \log \Tilde{\alpha}_\theta (T,U) \Tilde{\phi}_\theta (T,U).
\end{equation}


\section{Forward Variable Causal Compensation Estimator (FoCCE)}
\label{sec:FoCCE}
FoCC values $\gamma_\theta (\cdot,\cdot)$ defined in \eqref{eq:FoCC} are necessary for determining the actual likelihood in streaming transducers, though we could not find any analytic solution to access the probability terms in \eqref{eq:FoCC} based on the transducer network outputs $\Tilde{Y}_\theta (\cdot, \cdot)$ and $\Tilde{\phi}_\theta (\cdot, \cdot)$.
For this reason,
we propose to approximate $\gamma_\theta (\cdot,\cdot)$ with $\gamma_\omega (\cdot,\cdot)$ using a separate FoCC estimator (FoCCE) network parametrized by $\omega$.

Learning the probability ratio $\gamma_\omega (\cdot,\cdot)$ 
with a neural network is challenging, so we split it into two probability densities $\chi_\omega (\cdot,\cdot)$, $\Bar{\chi}_\omega (\cdot)$ and let the FoCCE network estimate them:
\begin{equation}
\label{eq:FoCCE_to_chi}
\gamma_\omega (t,u):= \Big{(} \frac{\chi_\omega (t,u)}{\Bar{\chi}_\omega (t)} \Big{)}^{\lambda_\gamma}
\end{equation}
where
\begin{equation}
\begin{split}
\label{eq:FoCCE_chi}
\chi_\omega (t,u):=&    P_\omega (\mathbf{x}_{e(t)+1:e(t+1)} |\mathbf{x}_{1:e(t)}, \mathbf{y}_{0:u}, z_u \leq t < z_{u+1} ), \\
\Bar{\chi}_\omega (t):=&    P_\omega (\mathbf{x}_{e(t)+1:e(t+1)} |\mathbf{x}_{1:e(t)})
\end{split}
\end{equation}
for $t \in [1,T), u \in [0,U]$, and $\lambda_\gamma$ is a non-negative scaling factor for FoCCE.
Intuitively, $\Bar{\chi}_\omega (\cdot)$ is a probability density of next-chunk input features given the current history of the input sequence (similar to autoregressive predictive coding~\cite{chung2019unsupervised_APC}), while the density $\chi_\omega (\cdot, \cdot)$ is also conditioned on the target sequence history.
With enough model capacity, it is assumed that both $\gamma_\theta (\cdot,\cdot)$ and $\gamma_\omega (\cdot,\cdot)$ converge to the true probability ratio $\gamma (\cdot,\cdot)$ as training progresses, therefore $\gamma_\theta (\cdot,\cdot) \approx \gamma_\omega (\cdot,\cdot)$ for well-trained models $\theta$ and $\omega$.

The probability densities $\chi_\omega (\cdot,\cdot)$, $\Bar{\chi}_\omega (\cdot)$ in \eqref{eq:FoCCE_chi} are estimated by the FoCCE network $\omega$, as illustrated in the middle of Figure~\ref{fig:pictorial_description}.
The architecture of this network is adapted from the conventional transducer networks, as shown below:
\begin{equation}
\label{eq:FoCCE_joiner_chi}
\begin{split}
\chi_\omega (t,u)=&\,   \mathrm{DensityEstimator}^\mathrm{\chi}_\omega ([f^{\chi}_t;g^{\chi}_u]),\\
\Bar{\chi}_\omega (t)=&\,   \mathrm{DensityEstimator}^\mathrm{\chi}_\omega ([f^{\chi}_t; \vec{0}]),    
\end{split}
\end{equation}
\begin{equation}
\label{eq:FoCCE_encoder_decoder}
\begin{split}
f^{\chi}_t =&\,    \mathrm{CausalEncoder}^\mathrm{\chi}_\omega (x_{1:e(t)}), \\
g^{\chi}_u =&\,    \mathrm{Predictor}^\mathrm{\chi}_\omega (y_{0:u}),    
\end{split}
\end{equation}
where $[\cdot;\cdot]$ denotes the concatenation, and $\vec{0}$ stands for a zero vector.
The $\mathrm{CausalEncoder}_\omega^{\chi} (\cdot)$ and $\mathrm{Predictor}_\omega^{\chi} (\cdot)$ respectively mean an encoder and a prediction network that function similarly to those of a transducer network.
The $\mathrm{CausalEncoder^\mathrm{\chi}_\omega (\cdot)}$ operates causally, just like the $\mathrm{CausalEncoder}_\theta (\cdot)$ described in \eqref{eq:streaming_encoder}, using the same context-end function $e(\cdot)$.
In contrast to the joiner of a transducer network, 
the $\mathrm{DensityEstimator}_\omega^{\chi} (\cdot)$ models probability densities in continuous space.
To model arbitrary densities, we utilized normalizing flows~\cite{rezende2015variational_normalizing_flows} to implement $\mathrm{DensityEstimator}_\omega^{\chi} (\cdot)$.

\begin{table*}[t]
 \caption{Word error rates (WERs) comparison of transducer models on the LibriSpeech and TED-LIUM3 datasets}
\label{tab:WERs}
\vspace{-0.25 in}
\begin{center}
\begin{small}
\renewcommand{\arraystretch}{1.1}
\begin{tabular}{ |l|c?c|c|c|c@rrrr|  }
 \hline
 \multirow{3}{*}{\begin{tabular}[c]{c} \textbf{Transducer model} \end{tabular}} & \textbf{Attention} & \multicolumn{3}{c|}{\textbf{FoCCE network hyperparam.}} & \textbf{\#} & \multicolumn{4}{c|}{\textbf{LibriSpeech WER [\%]} } \\ \cline{3-5} \cline{7-10}
 & \textbf{chunk}& \multirow{2}{*}{\begin{tabular}[c]{c} $\mathbf{\lambda}_\mathbf{\gamma}$\end{tabular}} & \multicolumn{2}{c|}{$\mathrm{\mathbf{CausalEncoder}}_{\mathbf{\omega}}^{\mathbf{\chi}}\mathbf{(\cdot)}$ \textbf{conv. module} } & \textbf{param.} & \multicolumn{2}{c}{\textbf{dev} } & \multicolumn{2}{|c|}{\textbf{test} } \\ \cline{4-5} \cline{7-10}
 &\textbf{size}&  & \textbf{\# module stacks} & \textbf{module dim.} & \textbf{(train)} & \textbf{clean} & \textbf{other} & \textbf{clean} & \textbf{other} \\
 \hline 
 Zipformer (non-streaming)    & full       & \multicolumn{3}{c|}{-} & 25.6M& 2.42 & 5.96  & 2.54 & 6.00 \\ \hline 
 Zipformer (streaming) &  \multirow{5}{*}{8}  & \multicolumn{3}{c|}{-} & 25.6M & 3.27 & 9.41  & 3.53 & 9.17  \\ \cline{3-10}
 $ $ + FoCCE (proposed) &       & 0.01 & 8 & 320 & 29.1M & 3.20  & 9.31  & 3.47 & 9.06 \\
 $ $ &       & 0.05 & " & " & " &\textbf{3.13} & 8.95  & \textbf{3.27} & \textbf{8.76} \\
 $ $  &    (160 ms)   & 0.25 & " & " & " & 3.32 & 9.40  & 3.60 & 9.20  \\
 $ $  &       & 0.05 & 4 & 256 & 27.3M & 3.26  &9.25  & 3.41 & 9.00  \\
 $ $  &       & " & 8 & 512 &33.2M & 3.14 & \textbf{8.90} & 3.34 & 8.78 \\
 \hline
 \hline
 \multirow{2}{*}{\begin{tabular}[c]{c} \textbf{Transducer model} \end{tabular}} & \textbf{Attention} & \multirow{2}{*}{\begin{tabular}[c]{c} $\lambda_{r}$ \end{tabular}} & \multicolumn{2}{c|}{$\mathbf{CausalEncoder}^{\chi}_{\omega}(\cdot)$ \textbf{conv. module}} & \textbf{\# param.}& \multicolumn{4}{c|}{\textbf{TED-LIUM3 WER [\%]}} \\
 \cline{4-5} \cline{7-10}
 & \textbf{chunk size} & & \textbf{\# module stacks} & \textbf{module dim.} & \textbf{(train)}& \multicolumn{2}{c|}{\textbf{dev}}& \multicolumn{2}{c|}{\textbf{test}}\\ \hline 
 Zipformer (non-streaming)    & full       & \multicolumn{3}{c|}{-} & 25.6M& \multicolumn{2}{c}{6.46} & \multicolumn{2}{c|}{5.91} \\ \hline
 Zipformer (streaming) &  \multirow{5}{*}{}  & \multicolumn{3}{c|}{-} & 25.6M & \multicolumn{2}{c}{9.43}  & \multicolumn{2}{c|}{8.57}  \\ \cline{3-10}
 $ $ + FoCCE (proposed) &     8  & 0.01 & 8 & 320 & 29.1M & \multicolumn{2}{c}{9.28}  & \multicolumn{2}{c|}{8.41} \\
 $ $ &    (160 ms)    & 0.05 & " & " & " &\multicolumn{2}{c}{\textbf{9.06}}  & \multicolumn{2}{c|}{\textbf{8.10}}\\
 $ $  &      & 0.25 & " & " & " & \multicolumn{2}{c}{9.35}  & \multicolumn{2}{c|}{8.51}  \\
 \hline
\end{tabular}
\vspace{-0.3 in}
\end{small}
\end{center}
\end{table*}

The FoCCE network $\omega$ is trained independently from the transducer model, maximizing the objective
\begin{equation}
\label{eq:FoCCE_loss}
L^{\chi}_\omega (\mathbf{x}_{1:T}, \mathbf{y}_{0:U}) :=
    \sum_{t=1}^{T-1}( \Bar{\chi}_\omega (t) + \frac{1}{U}\sum_{u=1}^{U} \chi_\omega (t,u) ).
\end{equation}
Based on the FoCC estimation $\gamma_\omega (\cdot,\cdot)$,
a streaming transducer network $\theta$ is trained to maximize the modified likelihood:
\begin{equation}
\label{eq:FoCCE_transducer_likelihood}
L^{mod}_{\theta,\omega} (\mathbf{x}_{1:T}, \mathbf{y}_{0:U}) :=
    \log \hat{\alpha}_{\theta,\omega} (T,U) \Tilde{\phi}_\theta (T,U),
\end{equation}
\begin{equation}
\begin{split}
\label{eq:streaming_forward_variable_and_transducer_recursion_FoCCE}
\hat{\alpha}_{{\theta,\omega}} (t,u) :=& \quad \:  \hat{\alpha}_{\theta,\omega} (t-1,u) \Tilde{\phi}_\theta (t-1,u) \mathrm{sg}\big{(}{\gamma_\omega (t-1,u)}\big{)}\\
    &+  \hat{\alpha}_{\theta,\omega} (t,u-1) \Tilde{Y}_\theta (t,u-1),
\end{split}
\end{equation}
with the initial and boundary values the same as $\alpha_\theta (t,u)$.
The modified likelihood in \eqref{eq:FoCCE_transducer_likelihood} approximates the actual likelihood in \eqref{eq:condition_likelihood_forward_streaming}, mitigating the deformed likelihood problem.
The right side of Figure~\ref{fig:pictorial_description} depicts the modified forward variable recursion rule described in \eqref{eq:streaming_forward_variable_and_transducer_recursion_FoCCE}.
In this rule, the stop-gradient operator $\mathrm{sg(\cdot)}$ is applied to $\gamma_\omega (\cdot,\cdot)$ to prevent the divergence of parameter values.
Note that from \eqref{eq:FoCCE_to_chi}, the modified streaming recursion in \eqref{eq:streaming_forward_variable_and_transducer_recursion_FoCCE} can be smoothly transitioned into the conventional recursion in \eqref{eq:transducer_recursion} by setting $\lambda_\gamma$ close to $0$.

The whole training objective is given by
\begin{equation}
\begin{split}
\label{eq:total_training_objective}
L^{tot}_{\theta,\omega} (\mathbf{x}_{1:T}, \mathbf{y}_{0:U}) :=& \; \lambda_{mod} L^{mod}_{\theta,\omega} (\mathbf{x}_{1:T}, \mathbf{y}_{0:U}) \\
    &+ \lambda{\chi}  L^{\chi}_{\omega} (\mathbf{x}_{1:T}, \mathbf{y}_{0:U}). 
\end{split}
\end{equation}
Both the transducer network $\theta$ and the FoCCE network $\omega$ are learned to maximize $\mathbb{E}_{\mathbf{x},\mathbf{y} \sim \mathbb{D}} L^{tot}_{\theta,\omega} (\mathbf{x},\mathbf{y})$ given a training set $\mathbb{D}$.
Note that the gradient backpropagated from the FoCCE network does not directly affect the transducer network. The transducer network is optimized only to maximize the modified likelihood.

\section{Experimental Results}
\label{sec:guidelines}

\subsection{Experimental Setting}
We followed the icefall~\cite{icefall_github} framework to train and evaluate the transducer models.

\subsubsection{Data Preparation}
We conducted experiments on the LibriSpeech~\cite{panayotov2015librispeech} and the TED-LIUM3~\cite{hernandez2018ted} datasets, with the designated training, validation, and evaluation sets.
We transformed speech waveforms into $80$-D log mel filterbank energies using a $25$ ms Hanning window with a $10$ ms stride, which were applied as input for the encoders.
The text data was encoded using a byte-pair encoding (BPE)~\cite{gage1994BPEorig, sennrich2015neural_BPE}, resulting in $500$ subword units, which were used as input for predictors.

\subsubsection{Neural Network Architecture}
For the transducer network, we employed the Zipformer~\cite{yao2023zipformer} from an existing recipe\footnote{\label{footnote:recipe}The Zipformer architecture and the training recipe can be found at \url{https://github.com/k2-fsa/icefall/blob/master/egs/librispeech/ASR/zipformer}. The \textbf{small-scaled model} recipe is used in this paper.}.
The Zipformer consists of an encoder with a $4\times$ total subsampling rate, a stateless prediction network \cite{ghodsi2020rnnt_stateless}, and a joint network followed by a softmax layer.
\mycomment{
The Zipformer transducer is composed of three main components:
\begin{itemize}
  \item Zipformer encoder: a $2\times$ time-strided convolutional front, followed by six attentional blocks and a $2\times$ time-subsampling layer, resulting in a $40$ ms stride of encoder representations.
  
  \item Stateless decoder: an embedding matrix followed by a 512-D 1D convolutional layer (context size of 2) with ReLU activation.
  
  \item Joiner: an embedding matrix, a 512-D dense layer with a tanh activation function, and a 500-D dense layer with softmax activation.
\end{itemize}

Within the Zipformer encoder, the attentional blocks incorporate a range of subsampling rates from 1 to 8, and each block contains a series of Zipformer small blocks, which consist of feedforward modules, self-attention layers, and convolution modules. For detailed technical information on these modules, we recommend referring to the recipe\footref{footnote:recipe}.
}
From the recipe, we made a few modifications to the Zipformer small block parameters: the number of small blocks in each block to $2$, feedforward dimension to $768$, encoder dimension to $256$, and encoder unmasked dimension to $192$. For the $\mathrm{CausalEncoder}_\theta(\cdot)$, we used an attentional block chunk size of $8$, resulting in $160$ ms of encoder latency.

The FoCCE network $\omega$ comprises three main components:
\begin{itemize}
  \item $\mathrm{CausalEncoder}^{\chi}_{\omega}(\cdot)$: eight stacks of chunk-wise causal convolution modules, which are identical to those in the Zipformer encoder\footref{footnote:recipe} but with a kernel size of 9.
  
  \item $\mathrm{Predictor}^{\chi}_{\omega}(\cdot)$: 
  a $128$-D LSTM layer.
  
  \item $\mathrm{DensityEstimator}^{\chi}_{\omega}(\cdot)$: masked autoregressive flow (MAF)~\cite{germain2015made} with a flow depth of $1$, two neural blocks per each depth, and a hidden dimension of $160$. 
\end{itemize}

To align the chunk boundaries of encoders of the FoCCE network and the transducer network,
we stacked the acoustic features into $4\times$ stacked features along the time-axis so that their dimension is $320$, which were processed by $\mathrm{CausalEncoder}^{\chi}_{\omega}(\cdot)$ to generate a causal context vector.
This vector, in combination with the output from $\mathrm{Predictor}^{\chi}_{\omega}(\cdot)$, acts as the condition for the MAF network $\mathrm{DensityEstimator}^{\chi}_{\omega}(\cdot)$;
Our model employs a standard Gaussian distribution as the prior.

\subsubsection{Training}
We trained the models for 40 epochs to ensure convergence.
We applied SpecAugment~\cite{park2019specaugment}\footref{footnote:recipe} to the acoustic features exclusively for the transducer encoder input.
We experimentally determined the FoCCE hyperparameters that minimized WERs such that $\lambda_{\gamma}=0.05$, $\lambda_{mod}=1$, and $\lambda_{\chi}=0.01$.
For all the experiments of streaming ASR, we used the identical parameter setting for the transducer network. Therefore all transducers have the same computational footprints at inference, and their performances are affected only by the likelihood estimation at training.

\subsubsection{Evaluation}
We assessed the performance of the transducer models in terms of word error rates (WER) using a beam search algorithm with a beam size of $4$.
We incorporated a left context of 256 frames while evaluating streaming models.

\subsection{Accuracy Improvement by FoCCE training}
Table~\ref{tab:WERs} displays the WERs for Zipformer transducers trained using different methods.
In LibriSpeech, FoCCE training on streaming transducers resulted in lower WERs, which amounts to $26.3$\% in test-clean and $12.3$\% in test-other of WER gaps between non-streaming and streaming baselines. FoCCE training also reduced the streaming and non-streaming WER gap by $17.7$\% in TED-LIUM3 test set.
The extent of WER improvement was sensitive to the hyperparameter $\lambda_\gamma$,
which arises from the fact that $\lambda_\gamma$ is calculated based on the division of two probability densities in continuous feature space, rather than two probabilities in discrete output space.

\section{Future Work}

We introduce FoCCE, the estimator for FoCC, that helps mitigate the deformed likelihood problem.
Our experiments show that FoCCE reduces WERs in streaming transducer training.
Future research should focus on refining the estimation process for FoCCE.

\bibliographystyle{IEEEtran.bst}
\bibliography{main}

\end{document}